\documentclass[12pt]{article}
\usepackage{epsfig}
\usepackage{amssymb}

\def\beq{\begin{equation}}
\def\eeq#1{\label{#1}\end{equation}}
\def\eeqn{\end{equation}}

\def\beqa{\begin{eqnarray}}
\def\eeqa#1{\label{#1}\end{eqnarray}}
\def\eeqan{\end{eqnarray}}
\def\CR{\nonumber \\ }

\let\bar=\overbar

\def\Dslash{\not{\hbox{\kern-4pt $D$}}}
\def\dslash{\not{\hbox{\kern-2pt $\del$}}}

\def\msb{{\bar{\ssstyle M \kern -1pt S}}}

\def\Title#1{\begin{center} {\Large {\bf #1} } \end{center}}

\begin{document}

\def\qslash{\not{\hbox{\kern-2pt $q$}}}

\Title{Effective description of QCD at high density}

\bigskip\bigskip


\begin{raggedright}

{\it Roberto Casalbuoni\index{Casalbuoni, R.}\\
Dipartimento di Fisica dell'Universita'\\
Sezione INFN di Firenze\\
I-50125 Firenze, ITALY}
\bigskip\bigskip
\end{raggedright}

\section{Introduction} Ideas about color superconductivity
go back to almost 25 years ago \cite{barrois}, but only recently
this phenomenon has received a lot of attention (for recent
reviews see refs. \cite{wilczek1,hsu}). The naive expectation is
that at very high density, due to the asymptotic freedom, quarks
would form a Fermi sphere of almost free fermions. However,
Bardeen, Cooper and Schrieffer proved that the Fermi surface of
free fermions is unstable in presence of an attractive, arbitrary
small, interaction.  Since in QCD  the gluon exchange in the
$\bar 3$ channel is attractive one expects the formation of a
coherent state of particle/hole  pairs (Cooper pairs). An easy
way to understand the origin of this instability is to remember
that the Fermi energy distribution for free fermions at zero
temperature is given by $f(E)=\theta(\mu-E)$ and therefore, the
maximum value of the energy (Fermi energy) is $E_F=\mu$. Then
consider the grand-potential   $F=E-\mu N$. Adding or subtracting
a particle (or adding a hole) to the Fermi surface does not
change $F$, since $F\to (E\pm E_F)-\mu(N\pm 1)=F$. We see that
the Fermi sphere of free fermions is higly degenerate. This is
the origin of the instability, because if we add to the Fermi
sphere  two particles bounded with a binding energy $E_B$, the
grand potential decreases, since $F-F_B=-E_B<0$. Therefore in
presence of an arbitrarily small attractive interaction, it is
energetically more favorable for fermions to pair and form
condensates.

We will discuss the methods for the quantitative evaluation of
the condensates via a toy model. We will present also the results
for the 2SC and CFL phases. Then we will introduce the effective
lagrangian describing the low energy degrees of freedom of the
CFL phase, that is the Goldstone bosons associated to the global
symmetry breaking. We will then show how to formulate in a
convenient way the idea of quasi-particles near the Fermi
surface, introducing a formalism similar to the one used in heavy
quark effective theory \cite{HQET}. Next we will couple the
quasi-particles to the Goldstone bosons in such a way to obtain
the right gap term in the fermionic lagrangian. Using a weak
coupling expansion, justified by asymptotic freedom, we will be
able to evaluate the self-energy of the Goldstone bosons. The
expansion at low momenta will allow us to get some of the
couplings appearing in the low-energy effective lagrangian. The
remaining couplings are then obtained through an analogous
evaluation of the gluon self-energy.

\section{Color condensation}

Thee physics of fermions at finite density and zero temperature
 can be treated in a systematic way by using  Landau's idea of
quasi-particles. An example is the Landau theory of Fermi
liquids. A  conductor is treated as a gas of almost free
electrons. However these electrons are dressed ones, where the
dressing takes into account the interactions which are neglected
in this description.  According to Polchinski
\cite{Polchinski:1992ed} this procedure just works because the
interactions can be integrated away in the usual sense of the
effective theories. Of course, this is a consequence of the
special nature of the Fermi surface, which is such that there are
practically no relevant or marginal interactions. In fact, all
the interactions are irrelevant except for the four-fermi
couplings between pairs of opposite momentum. Quantum corrections
make  the attractive ones relevant, and   the repulsive ones
irrelevant. This explains the instability of the Fermi surface of
free fermions against attractive four-fermi interactions, but we
would like to understand better the physics underlying the
formation of the condensates and how the idea of quasi-particles
comes about. To this purpose we will make use of a toy model
involving two Fermi oscillators describing, for instance, spin up
and spin down. Of course, in a finite-dimensional system there is
no spontaneous symmetry breaking, but this model is useful just
to illustrate the main points. We assume our dynamical system to
be described by the following Hamiltonian containing a quartic
coupling between the oscillators \beq H=\epsilon(a_1^\dagger
a_1+a_2^\dagger a_2)+Ga_1^\dagger a_2^\dagger a_1 a_2. \eeqn We
will study this model by using a variational principle. We start
introducing the following normalized trial wave-function
$|\Psi\rangle$\beq |\Psi\rangle=\left(\cos\theta+\sin\theta\,
a_1^\dagger a_2^\dagger\right)|0\rangle. \eeqn The di-fermion
operator, $a_1a_2$, has the following expectation value \beq
\Gamma\equiv\,\langle\Psi|\,a_1 a_2|\Psi\rangle =
-\sin\theta\cos\theta\label{gamma}. \eeqn Then we determine the
value of $\theta$ by looking for the minimum of the expectation
value of $H$ on the trial state \beq
\langle\Psi|H|\Psi\rangle=2\epsilon\sin^2\theta- G\Gamma^2. \eeqn
We get \beq 2\epsilon\sin 2\theta+2G\Gamma \cos 2\theta=0\,
\longrightarrow \, \tan 2\theta=-\frac{G\Gamma} \epsilon. \eeqn By
using the expression (\ref{gamma}) for $\Gamma$  we obtain the gap
equation \beq \Gamma=-\frac 1 2 \sin 2\theta=\frac 1 2
\frac{G\Gamma}{\sqrt{\epsilon^2+G^2\Gamma^2}}, \eeqn or \beq
1=\frac 1 2 \frac{G}{\sqrt{\epsilon^2+\Delta^2}},\eeqn where
$\Delta=G\Gamma$. Therefore the gap equation can be seen as the
equation determining the ground state of the system, since it
gives the value of the condensate. We can now introduce the idea
of quasi-particles in this particular context. Let us write the
hamiltonian $H$ as the sum of two pieces, a quadratic one, plus
another piece containing the interaction term defined in such a
way to have zero expectation value on the ground state
$|\Psi\rangle$. We have \beq H=H_0+H_{\rm res},\eeqn with \beq
H_0=\epsilon(a_1^\dagger a_1+a_2^\dagger a_2)-G\Gamma(a_1 a_2-
a_1^\dagger a_2^\dagger),\eeqn and \beq H_{\rm res}=G(a_1^\dagger
a_2^\dagger+\Gamma)\left(a_1 a_2-\Gamma\right),\eeqn where we
have neglected a constant term $ G\Gamma^2$. Let us now look for
a transformation on the Fermi oscillator variables such that
$H_0$ acquires a canonical form (Bogoliubov transformation). Such
a transformation is \beq
A_1=a_1\cos\theta-a_2^\dagger\sin\theta,~~~A_2=a_1^\dagger
\sin\theta +a_2\cos\theta,\eeqn and we get \beq H_0=
(\epsilon-\sqrt{\epsilon^2+\Delta^2})+\sqrt{\epsilon^2+\Delta^2}(A_1^\dagger
A_1+A_2^\dagger A_2).\eeqn Furthermore we can check that \beq
A_{1,2}|\Psi\rangle=0.\eeqn Therefore the operators $A_i^\dagger$
create out of the vacuum quasi-particles of energy \beq
E=\sqrt{\epsilon^2+\Delta^2}.\eeqn We see that the condensation
gives rise to the fermionic energy gap, $\Delta$. The Bogoliubov
transformation realizes the dressing of  the original operators
$a_i$ and $a_i^\dagger$ to the quasi-particle ones $A_i$ and
$A_i^\dagger$. Of course, the interaction is still present, but
part of it has been absorbed in the dressing process, and the
hope is that in such a way one is able to get a better starting
point for a perturbative expansion. As we have said this point of
view has been very fruitful in the Landau theory of conductors.

Let us  now discuss what have been done in order to determine the
gap in QCD at high density (see \cite{wilczek1}). At
asymptotically high-density one can, in principle, perform the
calculation starting from first principles, since QCD is weakly
coupled \cite{Son:1999uk} (for a more complete list of references
see \cite{wilczek1}). However the actual calculations are
unlikely to be extrapolated below a chemical potential of order
10$^8$ $GeV$ \cite{Rajagopal:2000rs}. Since the interesting
density regime for neutron stars and heavy ions is for
$\mu\lesssim 500~MeV$, one has to use phenomenological
interactions known to capture the essential features of QCD. In
the case of two massless flavors one can use the instanton vertex
producing a four-fermi interaction. For more flavors use has been
made of the one-gluon exchange approximation or, again, of a
four-fermi interaction with color, flavor, and spin structure
identical to the  one gluon exchange case. A practical way to
arrive at the gap equation is  to write down the Schwinger-Dyson
equation for the case at hand,  ignoring vertex corrections. One
gets an equation of the type \beq \Sigma(k)=-\frac{1}{2\pi^4}\int
d^4q\,G^{-1}(q)V(k-q), \eeqn where $G^{-1}(q)$ is the full
propagator and $V(p)$ is the vertex function, momentum
independent for a point-like interactions. This is an integral
equation for the self-energy which appears on the right-hand side
in $G^{-1}(q)$. Since one expects a diquark condensate it is
convenient to introduce the Nambu-Gorkov fields
\beq\Psi=\left(\matrix{\psi\cr\bar\psi^T}\right), \eeqn The wave
operator is then postulated of the form \beq G(q)=G(q)_{\rm
free}+\Sigma=\left(\matrix{\qslash+\mu\gamma_0 &
\gamma_0\Delta\gamma_0\cr \Delta &
(\qslash-\mu\gamma_0)^T}\right), \eeqn equivalent to introduce in
the lagrangian a mass gap term of the type $\psi^T C\Delta\psi$.
For a four-fermi interaction $\Delta$ is a matrix in
color-flavor-spin space. In the gluon-exchange case there is also
a momentum dependence. The typical gap equation for a four-fermi
interaction (neglecting antiparticle contribution) is
\beq\Delta=4G\int_0^\Lambda \frac{d^4 p}{(2\pi)^4}
\Big(\frac{\Delta}{p_0^2+ (|\vec p|-\mu)^2 +\Delta^2}\Big)
\label{18}.\eeqn When the interaction strength, $G\to 0$, and
$\Delta\ll\mu,\,\Lambda$, one finds \beq \Delta\propto \Delta
\mu^2\, G \log(\mu/\Delta)~{\to}~
\Delta\propto\mu\exp(-c/(G\mu^2)). \eeqn Replacing the effective
four-fermi interaction with one-gluon exchange  one expects \beq
\Delta\propto\mu\exp(-c/g^2). \eeqn But taking into account the
gluon propagator one gets a double-log contribution due to a
collinear infrared divergence arising from  the gluon propagator.
This divergence is regulated, for $q\to 0$, by the inverse of the
penetration length, which turns out to be of order $\Delta$
\cite{casa2}. In the weak coupling limit one gets
\cite{Son:1999uk} \beq \Delta\propto \Delta g^2
(\log(\mu/\Delta))^2~{\to}~ \Delta\propto\mu\exp(-c/g). \eeqn The
gap at large ${\mu}$ is much larger in QCD than in the case of a
point-like interaction. If the coupling $g$ is evaluated at $\mu$
and we  assume $1/g^2\approx \log\mu$ the exponential gives a very
weak suppression and for $\mu\to\infty$ we have $\Delta\to\infty$
and $\Delta/\mu\to 0$. Color superconductivity is bounded to
dominate physics at high density.

The phase structure of QCD at high density depends on the number
of flavors and there are two very interesting cases,
corresponding to two massless flavors (2SC) \cite{barrois,2SC}
and to three massless flavors (CFL) \cite{CFL1,CFL2}
respectively. In this talk we will be mainly concerned with the
latter case. The two cases correspond to very different patterns
of symmetry breaking. If we denote left- and right-handed quark
fields by $q_{iL(R)}^\alpha$ with $\alpha=1,2,3$, the $SU(3)_c$
color index, and $i=1,\cdots,N_f$ the flavor index ($N_f$ is the
number of massless flavors), in the 2SC phase the previous
calculations give  the following  color-flavor-spin structure
for the condensate
\beq \langle q_{iL(R)}^\alpha(\vec p\,)
C q_{jL(R)}^\beta(-\vec p\,)\rangle=\frac{\Delta}{G} \epsilon_{ij}
\epsilon^{\alpha\beta3}. \label{2SC} \eeqn Here $C=i\gamma^2\gamma^0$ is the
charge-conjugation matrix. The resulting symmetry
breaking pattern (barring  $U(1)$ factors) is \cite{barrois,2SC}
\beqa &SU(3)_c\otimes SU(2)_L\otimes SU(2)_R&\CR&\downarrow&\CR
&SU(2)_c\otimes SU(2)_L\otimes SU(2)_R& \eeqan The color group
 $SU(3)_c$ breaks down to the subgroup
$SU(2)_c$ but no global symmetry is broken. Although the baryon
number, $B$, is broken, there is a combination of $B$ and of the
broken color generator, $T_8$, which is unbroken in the 2SC phase.
Therefore no massless Goldstone bosons are present in this phase.
On the other hand, five gluon fields acquire mass whereas three
are left massless. It is worth to notice that  for the electric
charge the situation is very similar to the one for the baryon
number. Again a linear combination of the broken electric charge
and of the broken generator $T_8$ is unbroken in the 2SC phase.
The condensate (\ref{2SC})  gives rise to a gap, $\Delta$, for
quarks of color 1 and 2, whereas the two quarks of color 3 remain
un-gapped (massless). The resulting effective low-energy theory
has been described in \cite{sannino}.

Numerically,  by choosing a cutoff $\Lambda=800~MeV$ (see eq.
(\ref{18})), and fixing the coupling $G$ in such a way to
reproduce correctly the physics at zero density and temperature,
one finds \cite{wilczek1} \beqa
\mu&=&400~MeV~\Rightarrow~\Delta=106~MeV,\CR
\mu&=&500~MeV~\Rightarrow~\Delta=145~MeV. \eeqan

In this contribution we will be mainly interested in the
formulation of the effective theory for  three massless quarks.
An analysis similar to the previous one for this case leads to
the condensate \cite{CFL1,CFL2} \beq \langle{q^i_\alpha}_L(\vec
p\,)C{q^j_\beta}_L(-\vec p\,)\rangle=\frac{1}G
P^{ij}_{\alpha\beta}, \eeqn with \beq P^{ij}_{\alpha\beta}=\frac 1
3 (\Delta_8+\frac 1 8
\Delta_1)\delta^i_\alpha\delta^j_\beta+\frac 1
8\Delta_1\delta^i_\beta\delta^j_\alpha. \eeqn This originates the
breaking \beqa &SU(3)_c\otimes SU(3)_L\otimes SU(3)_R\otimes
U(1)_B\otimes U(1)_A&\CR &\downarrow&\\ &SU(3)_{c+L+R}\otimes
Z_2\otimes Z_2& \eeqan The $U(1)_A$ symmetry is broken at the
quantum level by the anomaly, but it gets restored at very high
density since the instanton contribution is suppressed
\cite{anomaly, small1,son1}. Again, for $\Lambda =800~MeV$, a
convenient choice for $G$ and $\mu=400~MeV$,
 one gets \cite{wilczek1}
\beq \Delta_8=80~MeV,~~~\Delta_1 =-176~MeV. \label{channel} \eeqn
The condensate can also be written in the form \beq \langle
q_{iL(R)}^\alpha(\vec p\,) C q_{jL(R)}^\beta(-\vec
p\,)\rangle\propto \epsilon^{ijX} \epsilon_{\alpha\beta
X}+\kappa(\delta^i_\alpha\delta^j_\beta+\delta_\beta^i\delta_\alpha^j)
\label{CFL}. \eeqn Notice that $\kappa=0$ corresponds to
$\Delta_1=-2\Delta_8$. Due to the Fermi statistics, the
condensate must be symmetric in color and flavor. As a
consequence the two terms appearing in eq. (\ref{CFL}) correspond
to the $(\mathbf{\bar 3},\mathbf{\bar 3})$ and
$(\mathbf{6},\mathbf{6})$ channels of $SU(3)_c\otimes
SU(3)_{L(R)}$. It turns out that $\kappa$ is small \cite{CFL1,
small1,small2} and therefore the condensation occurs mainly in
the $(\mathbf{\bar 3},\mathbf{\bar 3})$ channel (see also eq.
(\ref{channel})). The expression (\ref{CFL}) shows that the ground
state is left invariant by a simultaneous transformation of
$SU(3)_c$ and $SU(3)_{L(R)}$. This is called Color Flavor Locking
(CFL). The $Z_2$ symmetries arise since the condensate is left
invariant by a change of sign of the left- and/or right-handed
fields. As for the 2SC case the electric charge is broken but a
linear combination with the broken color generator $T_8$
annihilates the ground state. On the contrary the baryon number
is broken. Therefore there are $8+2$ broken global symmetries
giving rise to 10 Goldstone bosons. The one associated to
$U(1)_A$ gets massless only at very high density. The color group
is completely broken and all the gauge particles acquire mass.
Also all the fermions are gapped.

\section{Effective theory for the CFL phase}

We start introducing the Goldstone fields as the phases of the
condensates in the $(\mathbf{\bar 3},\mathbf{\bar 3})$ channel
\cite{casa1,hong1}\begin{equation} X_\alpha^i\approx
\epsilon^{ijk}\epsilon_{\alpha\beta\gamma}\langle q^j_{\beta
L}q^k_{\gamma L}\rangle^*,~~~ Y_\alpha^i\approx
\epsilon^{ijk}\epsilon_{\alpha\beta\gamma}\langle q^j_{\beta
L}q^k_{\gamma L}\rangle^*. \end{equation} Since quarks belong to
the representation $(\mathbf{3},\mathbf{3})$ of $SU(3)_c\otimes
SU(3)_{L(R)}$ and transform under $U(1)_B\otimes U(1)_A$
according to \beq qq_L\to e^{i(\alpha+\beta)} q_L,~~~q_R\to
e^{i(\alpha-\beta)} q_R,~~~e^{i\alpha}\in U(1)_B,~~~e^{i\beta}\in
U(1)_A, \eeqn the transformation properties of the fields $X$ and
$Y$ under the total symmetry group $G=SU(3)_c\otimes
SU(3)_L\otimes SU(3)_R\otimes U(1)_B\otimes U(1)_A$ are ($g_c\in
SU(3)_c$, $g_{L(R)}\in SU(3)_{L(R)}$) \beq X\to g_cXg_L^T
e^{-2i(\alpha+\beta)},~~~Y\to g_cYg_R^T e^{-2i(\alpha-\beta)}.
\eeqn The fields $X$ and $Y$ are $U(3)$ matrices and as such they
describe $9+9=18$ fields. Eight of these fields are eaten up by
the gauge bosons, producing eight massive gauge particles.
Therefore we get the right number of Goldstone bosons, $10 =18-8$.
These fields correspond to the breaking of the global symmetries
in $G$ (18 generators) to the symmetry group of the ground state
$H=SU(3)_{c+L+R}\otimes Z_2\otimes Z_2$ (8 generators). For the
following it is convenient to separate the $U(1)$ factors in $X$
and $Y$ in order to get new fields belonging to $SU(3)$ \beq
X=\hat X e^{2i(\phi+\theta)},~~~Y=\hat Y
e^{2i(\phi-\theta)},~~~\hat X,\hat Y\in SU(3). \eeqn The fields
$\phi$ and $\theta$ can be also described in an a more invariant
way through the determinants of $X$ and $Y$ \beq d_X={\rm
det}(X)=e^{6i(\phi+\theta)},~~~d_Y={\rm
det}(Y)=e^{6i(\phi-\theta)}, \eeqn with transformation properties
under $G$ \beq \hat X\to g_c\hat X g_L^T,~~~\hat Y\to g_c\hat Y
g_R^T,~~~\phi\to\phi-\alpha,~~~\theta\to\theta-\beta. \eeqn The
breaking of the global symmetry can  be discussed in terms of
gauge invariant fields  $d_X$, $d_Y$
and\beq\Sigma^i_j=\sum_\alpha (\hat Y_\alpha^j)^* \hat
X_\alpha^i\to \Sigma=\hat Y^\dagger \hat X. \eeqn The $\Sigma$
field describes the 8 Goldstone bosons corresponding to the
breaking of the chiral symmetry $SU(3)_L\otimes SU(3)_R$, as it
is made clear by the transformation properties of $\Sigma^T$,
\beq \Sigma^T\to g_L \Sigma^T g_R^\dagger.\eeqn That is $\Sigma^T$
transforms exactly as the usual chiral field. The other two
fields $d_X$ and $d_Y$ provide the remaining two Goldstone bosons
corresponding to the breaking of the $U(1)$ factors.

In order to build up an invariant lagrangian it is convenient to define
the following currents
\beq
J_X^\mu=\hat X D^\mu \hat X^\dagger=\hat X(\partial^\mu\hat X^\dagger+\hat
X^\dagger g^\mu),~~~J_Y^\mu=\hat Y D^\mu \hat Y^\dagger=\hat
Y(\partial^\mu\hat Y^\dagger+\hat Y^\dagger g^\mu),
\eeqn
with $g_\mu=ig_s g_\mu^a T^a/2$ the gluon field and $T^a=\lambda_a/2$ the
$SU(3)_c$ generators.
These currents have simple transformation properties under the full symmetry
group $G$, \beq J^\mu_{X,Y}\to g_c J^\mu_{X,Y}g_c^\dagger.\eeqn The most general
lagrangian, up to two derivative terms, invariant under $G$, the rotation
group $O(3)$ (Lorentz invariance is broken by the chemical potential term) and
the parity transformation defined as \beq\hat X(\vec x,t)\leftrightarrow
\hat Y(-\vec x,t),~~
\phi(\vec x,t)\to\phi(-\vec x,t),~~ \theta(\vec x,t)\to -\theta(-\vec x,t),~~
\eeqn
is \cite{casa1}
\beqa
&&{\cal L}=-\frac{F_T^2}4{\rm
Tr}\left[\left(J^0_X-J^0_Y)^2\right)\right]-\alpha_T\frac{F_T^2}4{\rm
Tr}\left[\left(J^0_X+J^0_Y)^2\right)\right]+\frac 12 (\partial_0\phi)^2+\frac
12(\partial_0\theta)^2\CR
&+&\frac{F_S^2}4{\rm Tr}\left[\left(\vec J_X-\vec
J_Y)^2\right)\right]+\alpha_S\frac{F_S^2}4{\rm
Tr}\left[\left(\vec J_X+\vec
J_Y)^2\right)\right]-\frac{v_\phi^2}2|\vec\nabla\phi|^2-
\frac{v_\theta^2}2|\vec\nabla\theta|^2.\label{lagrangian1}
 \eeqan Using $SU(3)_c$ color gauge invariance
we can choose $\hat X=\hat Y^\dagger$, making 8 of the Goldstone
bosons  disappear and giving mass to the gluons. The properly
normalized Goldstone bosons, $\Pi^a$,  are given in this gauge by \beq \hat
X=\hat Y^\dagger =e^{i\Pi^a T^a/F_T}. \eeqn Expanding eq.
(\ref{lagrangian1}) at the lowest order in the fields we get \beq
{\cal L}\approx\frac 12 (\partial_0\Pi^a)^2+\frac 12
(\partial_0\phi)^2+\frac 12(\partial_0\theta)^2 -\frac{v^2}
2|\vec\nabla\Pi^a|^2-\frac{v_\phi^2}2|\vec\nabla\phi|^2-
\frac{v_\theta^2}2|\vec\nabla\theta|^2,\eeqn with $v=F_s/F_T$. The
gluons $g_0^a$ and $g_i^a$ acquire Debye and Meissner masses
given by \beq m_D^2=\alpha_Tg_s^2 F_T^2,~~~m_M^2=\alpha_Sv^2g_s^2
F_T^2. \label{masses}\eeqn It should be stressed that these are not the true
rest
masses of the gluons, since there is a large wave function
renormalization effect making the gluon masses of the order of the
gap $\Delta$ (see later) \cite{casa2}. Since
this description is supposed to be valid at low energies (we
expect much below the gap $\Delta$), we could also decouple the
gluons solving their classical equations of motion neglecting
the kinetic term. The result from eq. (\ref{lagrangian1}) is \beq
g_\mu=-\frac 12 \left(\hat X\partial_\mu\hat X^\dagger+\hat
Y\partial_\mu\hat Y^\dagger\right). \eeqn It is easy to show that
substituting this expression in eq. (\ref{lagrangian1}) one gets \cite{casa2}
\beq {\cal L}=\frac{F_T^2}4\left({\rm
Tr}[\dot\Sigma\dot\Sigma^\dagger]-v^2{\rm
Tr}[\vec\nabla\Sigma\cdot\vec\nabla\Sigma^\dagger]\right)+ \frac
12\left(\dot\phi^2-v_\phi^2|\vec\nabla\phi|^2\right)+ \frac
12\left(\dot\theta^2-v_\phi^2|\vec\nabla\theta|^2\right). \eeqn
The first term is nothing but the chiral lagrangian
except for the breaking of the Lorentz invariance. This is a way of
seeing the quark-hadron continuity, that is the continuity
between the CFL and the nuclear matter in three flavor QCD. The
identification is perfect if one realizes that in nuclear matter
the pairing may occur in such a way to give rise to a superfluid
due to the breaking of the baryon number as it happens in CFL
\cite{schafer2}.

\section{Fermions near  the Fermi surface}

We will introduce now the formalism described in ref.
\cite{hong2} in order to evaluate several quantities of interest
appearing in the effective lagrangian. This formulation is based
on the observation that, at very high-density, the energy
spectrum of a massless fermion is described by states
$|\pm\rangle$ with energies  $E_\pm =-\mu\pm|\vec p|$ where
$\mu$  is the quark number chemical potential. For energies much
lower than the Fermi energy $\mu$, only the states $|+\rangle$
close to the Fermi surface. i.e. with  $|\vec p\,|\approx\mu$,
can be excited. On the contrary, the states $|-\rangle$ have
$E_-\approx -2\mu$ and therefore they decouple.

This can be seen more formally by writing the
four-momentum of the fermion as \beq p^\mu=\mu
v^\mu+\ell^\mu\label{momentum},\eeqn where $v^\mu=(0,\vec v_F)$, and $\vec v_F$
is the Fermi velocity defined as $\vec v_F=\partial E/\partial\vec p|_{\vec
p=\vec
p_F}$. For massless fermions $|\vec v_F|=1$. Since the hamiltonian for a
massless Dirac
fermion in a chemical potential $\mu$ is \beq
H=-\mu+\vec\alpha\cdot\vec
p\,,~~~\vec\alpha=\gamma_0\vec\gamma, \eeqn one has \beq
H=-\mu(1-\vec\alpha\cdot \vec v_F)+\vec \alpha\cdot\vec\ell.\eeqn
Then, it is convenient to introduce the projection operators \beq
P_\pm=\frac{1\pm\vec\alpha\cdot\vec v_F}2,\eeqn such that \beq
H|+\rangle=\vec\alpha\cdot\vec\ell\, |+\rangle,~~~
H|-\rangle=(-2\mu+\vec\alpha\cdot\vec\ell )\,|-\rangle.\eeqn

We can define
fields corresponding to the states $|\pm\rangle$ through the
decomposition \beq \psi(x)=\sum_{\vec v_F}\, e^{-i\mu v\cdot
x}\left[\psi_+(x)+\psi_-(x)\right],\eeqn where  an average over the
Fermi velocity $\vec v_F$ is
performed. The velocity-dependent fields $\psi_\pm (x)$ are given
by ($v^\mu=(0,\vec v_F)$)
 \beq \psi_\pm(x)=e^{i\mu v\cdot
x}\left(\frac{1\pm\vec\alpha\cdot\vec
v_F}2\right)\psi(x)~=~\int_{|\ell
|<\delta}\frac{d^4\ell}{(2\pi)^4} e^{-i\,\ell\cdot
x}\psi_\pm(\ell).\label{psimeno}\eeqn Since we are interested at
physics near the Fermi surface we integrate out all the modes
with $|\ell|>\delta$, where $\delta$ is a cut-off such that
$\delta\ll\mu$. At the same time we will choose $\delta$ greater than
the energy gap, $\Delta$. Since, as we shall see, physics around the Fermi
surface is effectively two-dimensional,  the results do not depend on the
value of $\delta$. Substituting inside the Dirac part of the QCD
lagrangian density one obtains ($V^\mu=(1,\vec v_f)$, $\tilde
V^\mu=(1,-\vec v_F)$)\beq {\cal L}=\sum_{\vec v_F}
\left[\psi_+^\dagger iV\cdot D\psi_++\psi_-^\dagger(2\mu+ i\tilde
V\cdot D)\psi_-+(\bar\psi_+i\Dslash_\perp\psi_- + {\rm
h.c.})\right],\eeqn where $\Dslash_\perp=D_\mu\gamma^\mu_\perp$ and
\beq
\gamma^\mu_\perp~=~\frac 1 2 \gamma_\nu\left(2 g^{\mu\nu}-
V^\mu\tilde V^\nu-\tilde V^\mu V^\nu \right).
\eeqn

We notice that the fields appearing in this
expression are evaluated at the same Fermi velocity because
off-diagonal terms are cancelled by the rapid oscillations of the
exponential factor in the $\mu\to\infty$ limit. This behaviour
can be referred to as the Fermi velocity superselection rule.

At the leading order in $1/\mu$ one has
 \beq iV\cdot
D\psi_+=0,~~~~\psi_-= -\frac{i}{2\mu}\gamma_0\Dslash_\perp\psi_+,\eeqn
showing
the decoupling of $\psi_-$ for $\mu\to\infty$. The
equation for $\psi_+$ shows also that only the energy and the
momentum parallel to the Fermi velocity are relevant variables in
the problem. We have an effective two-dimensional theory.

Eliminating the field $\psi_-$ we get \beq {\cal L}=\sum_{\vec
v_F}\left[\psi_+^\dagger iV\cdot D\psi_+-\frac 1{2\mu+i \tilde
V\cdot
D}\,\psi_+^\dagger(\Dslash_\perp)^2\psi_+\right].\label{effective}
\eeqn

The previous remarks apply to any theory describing massless
fermions at high density. The next step will be to couple this
theory in a $SU(3)_L\otimes SU(3)_R\otimes SU(3)_c$ invariant way
to  Nambu-Goldstone bosons (NGB)  describing the appropriate
breaking  for the CFL phase. Using a gradient expansion we
get an explicit  expression for  the decay coupling constant of
the Nambu-Goldstone bosons as well for their velocity.

\section{Goldstone bosons self-energy}

The invariant coupling between fermions and Goldstone fields
reproducing the symmetry breaking pattern of eq.
(\ref{CFL}) is proportional to \beq
\gamma_1\,Tr[\psi_L^T\hat X^\dagger]\,C\, Tr[\psi_L
\hat X^\dagger]+\gamma_2\,Tr[\psi_L^TC\hat X^\dagger\psi_L \hat X^\dagger]+{\rm
h.c.},\label{invariant}\eeqn and analogous relations for the
right-handed fields. Here the spinors are meant to be Dirac
spinors. The trace is operating over the group indices of the
spinors and of the Goldstone fields. Since the vacuum expectation
value of the Goldstone fields is $\langle \hat X\rangle=\langle
\hat Y\rangle=1$, we see that this coupling induces the right breaking
of the symmetry. In the following we will consider only the case
$\gamma_2=-\gamma_1\propto\Delta/2$, where $\Delta$ is the gap
parameter.

Since the transformation properties under the symmetry group of
the fields at fixed Fermi velocity do not differ from those of
the  quark fields, for both left-handed and right-handed fields
we get the effective lagrangian density \cite{casa2}\beqa {\cal
L}&=&\sum_{\vec v_F} \frac 1 2
\Big[\sum_{A=1}^9\left(\psi_+^{A\dagger}iV\cdot
D\psi_+^A+\psi_-^{A\dagger}i\tilde V\cdot
D\psi_-^A-{\Delta_A}\left({\psi_-^A}^TC\psi_+^A+{\rm
h.c.}\right)\right)\CR&-&{\Delta}\sum_{I=1,3}\left(Tr[(\psi_-
X_1^\dagger)^T C\epsilon_I(\psi_+
X_1^\dagger)\epsilon_I]+\rm{h.c.}\right)\Big],
\label{interaction}\eeqan where we have introduced the fields
$\psi_\pm^A$:
\beq\psi_\pm=\frac{1}{\sqrt{2}}\sum_{A=1}^9\lambda_A\psi^A_\pm.\eeqn
Here $\lambda_a~(a=1,...,8)$ are the Gell-Mann matrices normalized
as follows: $Tr(\lambda_a\lambda_b)=2\delta_{ab}$ and
$\lambda_9=\sqrt{ 2/ 3}~{\bf {1}}$. Furthermore
$\Delta_1=\cdots=\Delta_8=\Delta$, $\Delta_9=-2\Delta$, and
$X_1=\hat X-1$. Notice that  the NGB fields couple to fermionic
fields with opposite Fermi velocities. In this expression, as in
the following ones, the field $\psi_-$ is defined as $\psi_+$ with
$\vec v_F\to-\vec v_F$, and therefore it is not the same as the
one defined in (\ref{psimeno}).

The formalism becomes more compact by introducing
the Nambu-Gorkov fields \beq \chi=\left(\matrix{\psi_+\cr
C\psi^*_-}\right).\eeqn It is important to realize that the fields
$\chi$ and $\chi^\dagger$ are not independent variables. In fact,
since we integrate over all the Fermi surface, the fields
$\psi_-^*$ and $\psi_+$, appearing in $\chi$, appear also in
$\chi^\dagger$ when $\vec v_F\to -\vec v_F$. In order to avoid
this problem we can integrate over half of the Fermi surface, or,
taking into account the invariance under $\vec v_F\to -\vec v_F$,
we can simply integrate over all the sphere with a weight $1/8\pi$
instead of $1/4\pi$: \beq \sum_{\vec v_F}=\int\frac{d\vec
v_F}{8\pi}.\eeqn Then the first three terms in the lagrangian
density (\ref{interaction}) become \beq {\cal L}_0=\int\frac{d\vec
v_F}{8\pi}~ \frac 1 2\sum_{A=1}^9
\chi^{A\dagger}\left[\matrix{iV\cdot D & \Delta^A\cr\Delta^A
&i\tilde V\cdot D^*}\right]\chi^A,\label{kinetic}\eeqn so that, in
momentum space the free  fermion propagator is \beq
S_{AB}(p)=\frac{2\delta_{AB}}{V\cdot p\,\tilde V\cdot
p-\Delta_A^2}\left[\matrix{\tilde V\cdot p & -\Delta_A\cr-\Delta_A
& V\cdot p}\right].\eeqn

 We are now in position to evaluate the
self-energy of the Goldstone bosons through their couplings to the
fermions at the Fermi surface. There are two one-loop
contributions \cite{casa2}, one from the coupling $\Pi\chi\chi$
and a tadpole from the coupling $\Pi\Pi\chi\chi$ arising from eq.
(\ref{interaction}). The tadpole diagram contributes only to the
mass term and it is essential to cancel the external momentum
independent  term arising from the other diagram. Therefore, as
expected, the mass of the NGB's is zero. The contribution at the
second order in the momentum expansion is given by
 \beq
i\,  \frac{21-8\ln 2}{72\pi^2 F_T^2} \int\frac{d\vec
v_F}{4\pi}\sum_{a=1}^8 \Pi^a\,V\cdot p\, \tilde V\cdot p\,\Pi^a.
\eeqn Integrating over the velocities and going back to the
coordinate space we get \beq{\cal L}_{\rm eff}^{\rm kin}=
\frac{21-8\ln 2} {72\pi^2 F_T^2}\sum_{a=1}^8
\left(\dot\Pi^a\dot\Pi^a-\frac 1 3|\vec\nabla\Pi_a|^2\right).
\eeqn We can now determine the decay coupling constant $F_T$
through the requirement of getting the canonical normalization
for the kinetic term; this implies \beq F_T^2=\frac {\mu^2(21-8\ln
2)}{36\pi^2},\eeqn a result  obtained by many  authors
using different methods (for a complete list of the relevant
papers see the first reference of \cite{wilczek1}). We see also
that  $v^2=1/3$.  The origin of the pion velocity $1/\sqrt{3}$ is
a direct consequence of the integration over the Fermi velocity.
Therefore it is completely general and applies to all the NGB's
in the theory, including the ones associated to the breaking of
$U(1)_V$ and $U(1)_A$ ($v_\phi^2=v_\theta^2=1/3$); needless to
say, higher order terms in the expansion $1/\mu$ could change
this result.

The breaking of the Lorentz invariance exhibited by the pion
velocity different from one, can be seen also in the matrix
element $\langle 0|J_\mu^a|\Pi^b\rangle$. Its evaluation gives
\cite{casa2} \beq\langle 0|J_\mu^a|\Pi^b\rangle=iF_T\delta_{ab}
\tilde p_\mu, ~~~~\tilde p^\mu=(p^0, \vec p/3).\eeqn The current
is conserved, as a consequence of the dispersion relation
satisfied by the NGB's.

\section{Gluons self-energy}

The other couplings appearing in our effective lagrangian can be
obtained via a direct calculation of $m_D$ and $m_M$
\cite{casa2}. This is done evaluating the one-loop contribution
to the gluon self-energy. Also in this case there are two
contributions, one coming from the gauge coupling to the
fermions, whereas the other arises from the second term (sea-gull
like) appearing in the fermion effective lagrangian of eq.
(\ref{effective}). The results we find are \cite{casa2} \beq
m_D^2=g_s^2 F_T^2,~~~ m_M^2=\frac 1 3 m_D^2. \eeqn Comparison
with equation (\ref{masses}) shows that \beq \alpha_S=\alpha_T=1.
\eeqn

Performing a gradient expansion of the gluon self-energy one
finds that there is a wave function renormalization of order
$g_s\mu /\Delta \gg 1$ \cite{casa2}. In fact, considering  the
different components of the gluon field, the temporal, $g_0^a$
and the longitudinal, $g_L^{i\, a}$, and transverse, $g_T^{i\,
a}$  ones, defined as\beqa g_L^{i\, a}&=&\frac{\vec p\cdot \vec
g^a}{|\vec p|^2}\, p^i,\CR g_T^{i\, a}&=&g^{i\, a}-g_L^{i\,
a}~,\eeqan the following dispersion relations are obtained \beqa
g^{0a}:~~~3\alpha_1\,E^2-\alpha_1\,|\vec p\,|^2&=&m^2_D,\CR
g^{ia}_L:~~\alpha_1\,E^2-\alpha_2\,\frac{|\vec
p\,|^2}3&=&m_M^2,\CR g^{ia}_T:~~\alpha_1\,E^2-\alpha_3\,|\vec
p\,|^2&=&m_M^2,\eeqan where \beqa \alpha_1&=&\frac{\mu^2
g_s^2}{216\Delta^2\pi^2}\left(7+\frac{16}{3}\ln 2\right),~\CR
\alpha_2&=&-~\frac{\mu^2 g_s^2}{3240
\Delta^2\pi^2}\left(59-\frac{688}{3}\ln 2\right),~\CR
\alpha_3&=&-\frac{\mu^2
g_s^2}{3240\Delta^2\pi^2}\left(41-\frac{112}{3}\ln 2\right)~.
\eeqan The coefficients of the energy give just the square of the
wave function renormalization (notice that we have neglected the
term  coming from the free equation of motion, since the
renormalization part is much bigger than one). If we extrapolate
in momenta up to order $\Delta$ we see that the physical masses
of the gluons turn out to be of the order of the gap energy
($\approx 1.70\Delta$) \cite{casa2}. The validity of this
extrapolation has been recently confirmed in ref.
\cite{Gusynin:2001tt}. Wave function renormalization of order
$g_s\mu/\Delta$ for gauge fields in a dense fermionic  medium
appears to be a rather general phenomenon. For instance,
consider   the 2SC phase. The low energy degrees of freedom are 3
gluons and the almost free quarks of color 3. The symmetries
determining the effective lagrangian are: the gauge symmetry
$SU(2)_c$ and rotation invariance (Lorentz is broken being at
finite density). For the gluons one gets \cite{Rischke:2001cn}
\beq{\cal L}_{{\rm eff}}=\frac\epsilon 2\vec E^a\cdot\vec
E^a-\frac 1{2\lambda}\vec B^a\cdot\vec B^a, \eeqn with a
propagation velocity for the gluons given by $ v=1/
\sqrt{\epsilon\lambda}$\,. Values of $\epsilon$ and $\lambda$
different from 1 originate from wave function renormalization.
One finds \cite{Rischke:2001cn}\beq \epsilon=1+\frac{g_s^2\mu^2}
{18\pi^2\Delta^2}\approx\frac{g_s^2\mu^2}{18\pi^2\Delta^2},
~~~\lambda=1. \eeqn The strong coupling constant gets modified
\beq \alpha_s~\to~\alpha'_s=\frac{g^2_{\rm eff}}{4\pi
v}=\frac{g_s^2}{4\pi\sqrt{\epsilon}}=\frac{3}{2\sqrt{2}}
\frac{g_s\Delta}{\mu}, \eeqn due to the changes in the propagation
velocity and in the Coulomb force\beq g_s^2/r\to g_s^2/(\epsilon
r)~~ \Rightarrow~~ g_s^2\to g^2_{\rm eff}=g_s^2/\epsilon.\eeqn
Similar results hold for the massive gluons of type 4, 5, 6 and 7
which acquire a mass of order $\Delta$. Exceptions are the
spatial components (but not the time one) of the gluon 8. In this
case there is no wave function renormalization of the time
derivative and the mass is of order $g_s\mu$
\cite{Casalbuoni:2001ha}. Also the em dielectric constant gets
modified by the in-medium effects both in the CFL and in the 2SC
phases \cite{Litim:2001mv} \beq
\tilde\epsilon=1+\frac{r}{18\pi^2}\frac{\tilde
e^2\mu^2}{\Delta^2}, \eeqn where $\tilde e$ is the in-medium
rotated electric charge, and \beq r= 4 ~{\rm in~ CFL},~~ r=1~{\rm
in~ 2SC}. \eeqn

\begin{center}
{\bf Acknowledgments}
\end{center}\noindent
I would like to thank R. Gatto, M. Mannarelli  and  G. Nardulli for their
precious collaboration in the works originating this contribution.



\begin{thebibliography}{99}

\bibitem{barrois}

B. Barrois, Nucl.\ Phys.\ B {\bf 129}, 390 (1977);
S. Frautschi, {\it Proceedings of workshop on hadronic matter at
extreme density}, Erice 1978; D. Bailin and A. Love, Phys.\ Rep.\
{\bf 107} (1984) 325.

\bibitem{wilczek1}
K.~Rajagopal and F.~Wilczek,
hep-ph/0011333;

\bibitem{hsu}
S.D.H. Hsu,
hep-ph/0003140;
D.K. Hong, Acta Phys. Pol. {\bf B32} (2001) 1253, hep-ph/0101025;
M. Alford, hep-ph/0102047.

\bibitem{HQET}
N. Isgur and M.B. Wise, Phys. Lett. {\bf B232} (1989) 113 ;
{\it ibidem} Phys. Lett. {\bf B237} (1990) 527;
E. Eichten and B. Hill, Phys.
Lett. {\bf B234} (1990) 511;
H. Georgi, Phys. Lett. {\bf B240}
(1990) 447;
for a recent review see
A. V.  Manohar and M.B. Wise
{\it Heavy Quark Physics}, Cambridge University Press (2000).

\bibitem{Polchinski:1992ed}
J.~Polchinski,
Lectures presented at TASI 92,
hep-th/9210046.


\bibitem{Son:1999uk}
D.~T.~Son,
Phys.\ Rev.\ D {\bf 59} (1999) 094019, hep-ph/9812287.


\bibitem{Rajagopal:2000rs}
K.~Rajagopal and E.~Shuster,
Phys.\ Rev.\ D {\bf 62} (2000) 085007, hep-ph/0004074.

\bibitem{casa2}
R.~Casalbuoni, R.~Gatto and G.~Nardulli,
Phys.\ Lett.\ B {\bf 498} (2001) 179, hep-ph/0010321.


\bibitem{2SC}

M. Alford, K. Rajagopal and F. Wilczek, Phys.\ Lett.\ B {\bf
422}, 247 (1998), hep-ph/9711395;
R. Rapp, T. Sch\"afer, E.V. Shuryak and M. Velkovsky, Phys.\ Rev.\
Lett.\ {\bf 81},  53 (1998), hep-ph/9711396.

\bibitem{CFL1}

M. Alford, K. Rajagopal and F. Wilczek, Nucl.\ Phys.\ B {\bf 537},
 443 (1999), hep-ph/9804403.

\bibitem{CFL2}

T. Sch\"afer and F. Wilczek,  Phys.\ Rev.\ Lett.\ {\bf
82}, 3956 (1999), hep-ph/9811473.

\bibitem{sannino}

R. Casalbuoni,  Z. Duan and F. Sannino, Phys.\ Rev.\ D {\bf 62},
 094004 (2000), hep-ph/0004207; 
{\it ibidem}  D {\bf 63},
114026 (2001), hep-ph/0011394.


\bibitem{anomaly}

R. Rapp, T. Sch\"afer , E.V. Shuryak and M. Velkovsky, Ann.\ of
Phys.\ {\bf 280},  35 (2000),  hep-ph/9904353.


\bibitem{small1}

T. Sch\"afer, Nucl.\ Phys.\ B {\bf 575},  269 (2000),
hep-ph/9909574.





\bibitem{son1}

D.T. Son and M.A. Stephanov, Phys.\ Rev.\ D {\bf 61}, 074012
(2000), hep-ph/9910491;
{\it ibidem} Erratum D {\bf 62},
059902 (2000), hep-ph/0004095.

\bibitem{small2}

I.A. Shovkovy and L.C. Wijewardhana, Phys.\ Lett.\ B {\bf 470},
 189 (1999), hep-ph/9910225.


\bibitem{casa1}

R. Casalbuoni and R. Gatto, Phys.\ Lett.\ B {\bf 464},  111
(1999) ,
hep-ph/9908227.

\bibitem{hong1}

D.K. Hong, M. Rho and I. Zahed, Phys.\ Lett.\ B  {\bf 468}, 261
(1999),
hep-ph/9906551.



\bibitem{schafer2}

T.Sch\"afer and F. Wilczek, Phys.\ Rev.\ Lett.\ {\bf 82}, 3956
(1999), hep-ph/9811473.

\bibitem{hong2}

D.K.  Hong,  Phys.\ Lett.\ B {\bf 473}, 118 (2000),
hep-ph/9812510; 
D.K.  Hong {\it Nuclear Physics} B {\bf 582}, 451 (2000),
hep-ph/9905523; 
S.R.  Beane, P.F.  Bedaque and M.J.  Savage,  Phys.\ Lett.\ B {\bf
483},  131 (2000),  hep-ph/0002209.

\bibitem{Gusynin:2001tt}
V.~P.~Gusynin and I.~A.~Shovkovy,
hep-ph/0108175.


\bibitem{Rischke:2001cn}

D.~H.~Rischke, D.~T.~Son and M.~A.~Stephanov,
Phys.\ Rev.\ Lett.\  {\bf 87} (2001) 062001, hep-ph/0011379.


\bibitem{Casalbuoni:2001ha}

R.~Casalbuoni, R.~Gatto, M.~Mannarelli and G.~Nardulli,
hep-ph/0107024.


\bibitem{Litim:2001mv}

D.~F.~Litim and C.~Manuel,
hep-ph/0105165.
\end{thebibliography}
\end{document}